\documentclass[showpacs,amsmath,amssymb]{revtex4}
\usepackage{amssymb}
\usepackage[dvips]{graphicx}

\begin{document}

\title{Upper critical field in the model with finite range interaction between electrons}
\author{A. V. Shumilin$^1$, V.V. Baranov$^2$ and V.V. Kabanov$^3$}

\affiliation{$^1$ Ioffe Institute, 194021 St.-Petersburg, Russia}
\affiliation{$^2$ Department of Physics, University of Antwerp, Groenenborgerlaan 171, 2020
Antwerp, Belgium.}
\affiliation{$^3$ Department for Complex Matter, Jozef Stefan Institute, 1001 Ljubljana, Slovenia}

\begin{abstract}
We develop a theory of the upper critical field in a BCS superconductor with
a nonlocal interaction between electrons.
We have shown that the nonlocal interaction is characterized by the parameter $k_F\rho_0$ where
$k_F$ is the Fermi momentum and $\rho_0$ is the radius of electron-electron interaction. The
presence of the external magnetic field leads to the generation of additional components of the
order parameter with different angular momentum. This effect leads to the enhancement of the
upper critical field above the orbital limiting field. In addition the upward curvature in the temperature dependence of $H_{c2}(T)$ in the clean limit is predicted.
The impurity scattering suppresses the effect in the dirty limit.
\end{abstract}


\maketitle
\section{Introduction}\label{sec1}
The upper critical field $H_{c2}$ is one of the important characteristics of type-II
superconductors. When the field is sufficiently high, the superconductivity is destroyed and the
field uniformly penetrates to the sample. With the continuous decrease of the field
superconducting regions start to nucleate spontaneously at the upper critical filed
$H=H_{c2}(T)$. Since within these regions the order parameter is small the linearized
phenomenological Ginzburg-Landau equations are applicable in the vicinity of zero field $T_c$.
This leads to linear in $T$ upper critical field $H_{c2}(T)\propto T_c-T$\cite{deGennes}. At
zero temperature, $H_{c2}(0)$ is normally below the Clogston-Chandrasekhar \cite{clogston} or the Pauli pair-breaking limit given by $H_p = 1.84T_c$ (in tesla) for the singlet pairing.

Recent high-magnetic-field studies in cuprates \cite{Mackenzie,Osofsky,Ando,Gantmakher}, $MgB_2$,
$Ba_{1-x}K_xBiO_3$ and pnictide superconductors \cite{Lyard,Escribe,Mun}, spin-ladders
\cite{Nakanishi} and organic superconductors \cite{Chaikin}  have revealed a non-BCS upward curvature of the resistive $H_{c2}(T)$. In some cases \cite{Chaikin,Lawrie,Walker} the Pauli
limit was exceeded by several times. A non-linear temperature dependence in the vicinity of Tc
has been unambiguously observed in a few samples \cite{Gantmakher,Lyard,Escribe,Lawrie,Alex,Shulga}.
The observation of the departure from the BCS behaviour creates some controversy in the
interpretation of the resistive critical field \cite{Wen,Morozov,Grissonnanche}. If in some cases
there are little doubts that the resistive transition corresponds to the upper critical field $H_{c2}$ \cite{Gantmakher,Lyard,Escribe,Nakanishi,Chaikin,Lawrie,Alex,Shulga}, some measurements
on high temperature superconductors suggest that the real upper critical field is much higher than
the resistive transition \cite{Wen,Morozov,Grissonnanche}.
Indeed, the thermodynamic determination of $H_{c2}$\cite{Grissonnanche,Carrington,Roulin} and
anomalous diamagnetism above the resistive transition \cite{Wen,Junod} seem to justify such a
conclusion.

Several theoretical concepts have been proposed in order to explain a non-BCS upward curvature of
the resistive $H_{c2}(T)$. Some of the concepts are based on the fact that the size of the
pairs is smaller than the average distance between electrons and therefore superconducting
state may be approximated by the weakly interacting charged Bose gas\cite{Alex,Zavaritsky}.
In that case upper critical field has universal temperature dependence $(T_c-T)^{3/2}$
near $T_c$ \cite{Alex,Zavaritsky}.

Another approach is based on the multiband picture \cite{Lyard,Shulga}. In the case when
the Fermi surface has few sheets the gaps on different branches of the Fermi surface are
independent functions with different coupling constants. In such a system the upper critical
field may deviate considerably from the classical BCS behaviour leading to the weak upward
curvature in the temperature dependence of $H_{c2}$ \cite{Lyard,Shulga}.

Similar temperature dependence of the upper critical field may be caused by the field induced
mixture of the superconducting gaps of different symmetries\cite{Joynt,Balatsky,Kabanov}. On the basis
of the phenomenological Ginzburg-Landau approach it has been suggested that in addition to the d-wave
gap the external field may lead to generation of the s-wave\cite{Joynt}, additional d-wave\cite{Balatsky} or p-wave\cite{Kabanov} components of the order parameter. This effect is caused by the fact that symmetry allows
nontrivial gradient terms in the free energy\cite{Joynt,Balatsky} including Lifshitz
invariants\cite{Kabanov}. This type of coupling may also lead to unusual temperature dependence
of the upper critical field $H_{c2}(T)$\cite{Joynt,Kabanov}. This type of
Ginzburg-Landau equations were later derived from Gorkov equations assuming nonlocal potential between
electrons\cite{Yong}. Note that the approach based on Ginzburg-Landau equation is restricted to the relatively high
temperatures $T_c - T \ll T_c$.

Here we generalize Werthamer-Helfand-Hohenberg (WHH) theory
\cite{Werthamer1,Werthamer2,Werthamer3} to the case of a nonlocal interaction between electrons.
It allows us to go beyond the high-temperature limit and consider the upper critical field behavior caused by the intermixture of the different order parameters at arbitrary temperature. We demonstrate that the nonlocal interaction is characterized by the parameter $k_F\rho_0$ where
$k_F$ is the Fermi momentum and $\rho_0$ is the radius of electron-electron interaction. We show
that in the presence of the external magnetic field the finite radius of electron-electron
interaction leads to the generation of additional components of the order parameter with
different angular dependence. This effect leads to the enhancement of the upper critical field
above the orbital limiting field as
well as to the upward curvature in the temperature dependence of $H_{c2}(T)$ in the clean limit.
The increase of the impurity concentration suppress the effect.

\section{Main equations}\label{sec2}

We consider superconductor with a non-local pairing potential $V({\mathbf r}-{\mathbf r'})$ that explicitly
depends on coordinates ${\mathbf r},{\mathbf r'}$ and can be characterized by the radius of interaction $\rho_0$. The superconductor can be described with the
Hamiltonian
\begin{eqnarray}
H&=&\sum_{\sigma=\uparrow,\downarrow}\int d{\mathbf r}\Psi_{\sigma}^{\dag}({\mathbf r})\hat{\xi} \Psi_{\sigma}({\mathbf r})\cr
&+&\int d{\mathbf r}d{\mathbf r'}V({\mathbf r}-{\mathbf r'}) \Psi_{\uparrow}^{\dag}({\mathbf r})\Psi_{\downarrow}^{\dag}({\mathbf r'}) \Psi_{\downarrow}({\mathbf r'})\Psi_{\uparrow}({\mathbf r}).
\label{hamiltonian}
\end{eqnarray}
Here $\hat{\xi}={1\over{2m^*}}(i{\mathbf \nabla}+e{\mathbf A})^2-\mu$, $m^*$ is the effective mass of electron, $\mu$ is the chemical potential, and ${\mathbf A}$ is the vector potential.

The equation of motion for the field operator $\Psi_{\uparrow}({\mathbf r},t)$ derived from the Hamiltonian (\ref{hamiltonian}) is:
\begin{equation}\label{motion1}
\left({\partial\over{\partial t}}+\hat{\xi} \right)\Psi_{\uparrow}({\mathbf r},t)=-\int d{\mathbf r''} \Delta({\mathbf r}, {\mathbf r''})\bar{\Psi}_{\downarrow}({\mathbf r''},t)
\end{equation}
where $\Delta({\mathbf r}, {\mathbf r'})=V({\mathbf r}-{\mathbf r'})\langle\Psi_{\downarrow}(
{\mathbf r'})\Psi_{\uparrow}({\mathbf r})\rangle$ is the order parameter of the superconductor.
In contrast to the BCS model the order parameter explicitly depends on two coordinates ${\bf r}$ and ${\bf r}'$.

Multiplying equation (\ref{motion1}) to $\Psi_{\downarrow}({\mathbf r'},t')$ we obtain the equation for the
anomalous Green's function $F({\mathbf r} t,{\mathbf r'}t')$. We write this equation in terms of Matsubara frequencies $\omega_n=\pi T(2n+1)$, where $n$ is integer
\begin{equation}\label{eq-F}
(-i\omega_n+\hat{\xi})F_{\omega_n}({\mathbf r},{\mathbf r'})=\int d{\mathbf r''} \Delta({\mathbf r}, {\mathbf r''})G_{-\omega_n}({\mathbf r'},{\mathbf r''}).
\end{equation}
The anomalous Green's function is closely related to the order parameter
\begin{equation}\label{def-Del}
\Delta({\mathbf r},{\mathbf r'})=V({\mathbf r}-{\mathbf r'})T\sum_{\omega_n} F_{\omega_n}({\mathbf r}, {\mathbf r'}).
\end{equation}
In the present study we are interested in the upper critical field. The order parameter at magnetic field close to critical is small. It
allows us to use a normal state expression for the normal Green's function $G_{\omega_n}({\mathbf r},{\mathbf r'})$ \cite{AGD}.

Expressions (\ref{eq-F},\ref{def-Del}) allow  us to give a closed equation for the order parameter in terms of the normal state Green's functions
\begin{eqnarray}
&&\Delta({\boldsymbol \rho},{\mathbf R})=-TV({\boldsymbol \rho})\sum_n\int  G_{\omega_n}({\mathbf R}+{\boldsymbol \rho/2},{\mathbf R'}+{\boldsymbol \rho'/2}) \cr
&&\Delta({\boldsymbol \rho'},{\mathbf R'})G_{-\omega_n}({\mathbf R}-{\boldsymbol \rho/2},{\mathbf R'}-{\boldsymbol \rho'/2}) d{\boldsymbol \rho'} d{\mathbf R'}.
\label{delta2}
\end{eqnarray}
Here we introduce a new set of variables ${\bf R}$ and ${\boldsymbol \rho}$ for the order parameter $\Delta$ and the pairing potential $V$. This variables are related to the variables ${\bf r}$ and ${\bf r}'$ used in Eq.~(\ref{def-Del}) as follows.
The variable ${\bf R} = ({\bf r} + {\bf r}')/2$ describes the motion of the center of mass of the Cooper pair. It is related to the macroscopic distribution of the order parameter in the sample. At zero magnetic field the order parameter is uniform and does not depend on ${\bf R}$.
The variable ${\boldsymbol \rho}={\mathbf r}-{\mathbf r'}$ describes the relative motion of electrons in the Cooper pair. It appears due to the non-local pairing potential and describes the symmetry of the order parameter. If we select local pairing potential
$V \propto \delta({\boldsymbol \rho})$, ${\boldsymbol \rho}$ should be always equal to zero. Our theory is reduced to the conventional BCS theory in this case. In the more general case the important values of ${\boldsymbol \rho}$ are of the order of $\rho_0$.

The internal coordinate ${\boldsymbol \rho}$ allows an additional degree of freedom to the superconductivity. We are especially interested in the angular dependance of ${\boldsymbol \rho}$. We will show that the order parameter can be divided into the components related to the different angular momenta of ${\boldsymbol \rho}$. These components are independent without the external magnetic field. In the magnetic field these components are intermixed. It can lead to the upward curvature of $H_{c2}(T)$ dependence. In the present study we focus on the 2D case. It may be realized experimentally in the atomically thin films
or in very anisotropic superconductors similar to high-$T_c$ superconductors. In this case the components of the order parameter can be classified by the projection of the angular momentum on the axis perpendicular to the 2D plane. In principle a similar theory can be formulated in 3D when the components of the order parameter are classified by the value of the angular momentum. However this issue goes beyond of the present study.

\section{Clean limit}\label{sec3}

We start our consideration from the clean limit ($l>>\xi$) where the mean free path $l$ is much larger that the coherence length $\xi$ of the
superconductor.
In this case we neglect the effects of the impurities and we can give an explicit expression for the Green's function.
\begin{equation}
G_{\omega_n}^{(0)}({\mathbf R})=
\begin{cases}
{-im^*\over{\sqrt{2\pi k_F R}}}e^{i(k_F R-\pi/4)-{|\omega_n|\over{v_F}}R},& \omega_n>0, \\
{im^*\over{\sqrt{2\pi k_F R}}}e^{-i(k_F R-\pi/4)-{|\omega_n|\over{v_F}}R},& \omega_n<0.
\end{cases}
\label{GF}
\end{equation}
Here we consider  the limit $k_FR \gg 1$. The expression (\ref{GF}) corresponds to the normal Green's function without magnetic field. The Green's function
in the magnetic field is related to $G_{\omega_n}^{(0)}$ as $G_{\omega_n}({\mathbf R}, {\mathbf R}') = G_{\omega_n}^{(0)}({\mathbf R}'-{\mathbf R})\exp(i\phi({\mathbf R},{\mathbf R'}))$ where $\phi({\mathbf R},{\mathbf R'})\approx e\int_{\mathbf R}^{\mathbf R'}d{\mathbf s}{\mathbf A} ({\mathbf s})$. We also neglect the paramagnetic effects assuming that they are small as $\Delta/E_F \ll 1$. These effects may be important in the case where the upper critical field reaches Pauli pair-breaking limit. This usually occurs due to strong impurity scattering. However, as it will be shown later, strong impurity scattering suppresses the field induced mixture of pairing with different angular momenta.

The equation for the order parameter with the Green's functions (\ref{GF}) is
\begin{eqnarray} \label{int-eq-cl}
\Delta({\boldsymbol \rho},{\mathbf R})&=&-V({\boldsymbol \rho})\int d{\bf R'}d{\boldsymbol \rho'}{\cal K}_0(\widetilde{R})\Delta({\boldsymbol \rho'}, {\mathbf R'})  \cr
&&\exp\bigl(2i\phi({\mathbf R},{\mathbf R'})\bigr)\cos\left(k_F\widetilde{\bf R}{\boldsymbol \varrho}\over{\widetilde{R}}\right)
\end{eqnarray}
where $\widetilde{\bf R}={\mathbf R'}-{\mathbf R}$, $\widetilde{R} = |\widetilde{\bf R}|$,
${\boldsymbol \varrho}={\boldsymbol \rho'}-{\boldsymbol \rho}$
and the kernel
\begin{equation}
{\cal K}_0(R)={m^{*2}T\over{2\pi k_F R \sinh{(2\pi TR/v_F)}}}.
\end{equation}

Let us assume for simplicity that the pairing potential has the form
$V({\boldsymbol \rho})=-V_0\delta(|{\boldsymbol \rho}|-\rho_0)$, where $\rho_0$ plays the role of
the interaction radius. The potential with the finite range of interaction leads to the
formation of pairs with different angular momenta. Indeed, Fourier components of the potential
$V({\mathbf k'}-{\mathbf k})=V(\varphi',\varphi)$, where $k,k'=k_F$, depend only on polar angles $\varphi,\varphi'$ of vectors ${\mathbf k},{\mathbf k'}$. Calculating the matrix elements
$V_{n,n'}=\int d\varphi d\varphi' V(\varphi',\varphi)e^{in\varphi}e^{-in'\varphi'}$ we obtain:
\begin{equation}
V_{n,n'}=-V_0\rho_0J_n^2(k_F\rho_0)\delta_{n,n'}.
\end{equation}
Here $J_n(x)$ is the Bessel function. Therefore this potential leads to the pairing in the
channels with nonzero orbital moments $n$ with the critical temperatures
$T_{cn}=T_{c0}\exp(\lambda_0^{-1}-\lambda_n^{-1})$ and
$\lambda_n={m^*\over{2\pi}}V_0\rho_0J_n^2(k_F\rho_0)$. Note that the strength of the pairing is determined by
the parameter $k_F\rho_0$. If $k_F\rho_0<<1$ all channels except the $n=0$ channel are suppressed
because $J_n(x)\approx x^n$ when $x<<1$. When $k_F\rho_0 \gtrsim 1$ one of the channels with $n\ne 0$ can
have the largest $T_c$ and corresponds to the main order parameter of the superconductor \cite{Mineev}.
In real systems the smallest radius of interaction is determined by
the screening radius and therefore the situation $k_F\rho_0>1$ seems to be natural. Similar to the conventional BCS theory we do not try to choose a realistic pairing potential. Real interaction between electrons is complicated and may not be characterized by the potential. On the other hand we believe that the exact form potential is not very important for our results. All the final results will be expressed in terms of critical temperatures $T_{cn}$ in different pairing channels (see Eqs.(19),(21) and (29)). Therefore the exact form of the potential drops out from our results. Also we want to note that as long as the pairing potential acts only on the electrons near Fermi surface it can be reduced to its dependence on angles $\varphi$, $\varphi'$. This dependence define $V_{n,n'}$ in the Eq. (9) and the critical temperatures $T_{cn}$. We believe that any coordinate form of the potential that results in the same temperatures $T_{cn}$ should lead to similar results for $H_{c2}$. Here we choose the potential to make the calculations as simple as possible.

For the discussed choice of the pairing potential the order parameter can be written as:
\begin{equation}
\Delta({\boldsymbol \rho},{\mathbf R})=\Delta(\rho_0,\varphi,{\mathbf R})=\sum_n\Delta_n({\mathbf R})e^{in\varphi}.
\end{equation}
Substituting this expansion back to the integral equation (\ref{int-eq-cl}) we obtain:
\begin{eqnarray}
\Delta_n({\mathbf R})&=&V_0\rho_0\sum_{n'}\int d{\mathbf R'}{\cal K}_0(\widetilde{ R})L_{nn'}(\widetilde{\bf R}) \cr
&&\exp(2i\phi({\mathbf R},{\mathbf R'}))\Delta_{n'}({\mathbf R'}),
\end{eqnarray}
where the  matrix $L_{nn'}$ is defined as:
\begin{equation}
L_{n,n'}=e^{i(n'-n)\theta}
\begin{cases}
(-1)^{l+k}J_{2l}(k_F\rho_0)J_{2k}(k_F\rho_0), \\|n|=2l,|n'|=2k \\
(-1)^{l+k}J_{2l+1}(k_F\rho_0)J_{2k+1}(k_F\rho_0), \\|n|=2l+1,|n'|=2k+1
\end{cases}
\label{Lmm}
\end{equation}
where $\theta$ is the polar angle of the vector $\widetilde{\bf R}$.

The matrix $L_{n,n'}$ connects the
angular momenta of the internal coordinate ${\boldsymbol \rho}$ and the averaged coordinate ${\bf R}$. While the total momentum is conserved,
the non-diagonal matrix elements of $L_{n,n'}$ allow the transfer of the momentum between the degrees of freedom corresponding to ${\boldsymbol \rho}$
and ${\bf R}$.

Following the procedure described in Ref.
\cite{Werthamer2} we expand $\Delta({\bf R}')$ into the series over ${\bf R}' - {\bf R}$ and join the space derivatives with
vector potential into the single operator ${\bf D} ={\partial\over{\partial{\mathbf R}}}+2ie{\mathbf A} $:
\begin{eqnarray} \label{Del-mat-1}
\Delta_n({\mathbf R})&=&V_0\rho_0\sum_{n'}\int d\widetilde{\bf R}{\cal K}_0(\widetilde{R})L_{nn'}(\widetilde{\bf R}) \cr
&&\exp{({\mathbf D} \cdot \widetilde{\bf R})}\Delta_{n'}({\mathbf R}).
\end{eqnarray}
Note that the operator ${\mathbf D}$ acts only on the
coordinate ${\mathbf R}$.

Let us choose the gauge ${\mathbf A}=(0,Hx,0)$. With this gauge the order parameters $\Delta_n$ are not
dependent on $y$ and can be considered as functions $\Delta_n (x)$. Moreover, this gauge allows us to
relate different terms $\Delta_n$ of the order parameter to different coordinate functions $\psi_m$
\begin{equation} \label{psi-osc}
\psi_m = \left( \frac{1}{\pi \hbar L_H^2} \right)^{1/4} \frac{1}{\sqrt{2^m m!}}e^{-x^2/2L_H^2} H_m \left( \frac{x}{L_H} \right)
\end{equation}
that correspond
to the eigenfunctions of the harmonic oscillator. Here $L_H^{-1}=\sqrt{{2\pi H\over{\phi_0}}}$ is the magnetic length, $\phi_0$ is the flux quanta.
$H_n$ are the Hermite polynomials.

To apply the basis (\ref{psi-osc}) it is useful to make the expansion
${\mathbf D}\widetilde{\bf R}={\widetilde{R}\over{2}}(D_+e^{-i\theta}+D_-e^{i\theta})$
where $D_{\pm}=D_x \pm iD_y$, $\theta$ is the polar angle of the vector $\widetilde{\bf R}$. The action of the operators $D_-$ and $D_+$
on the functions $\Psi_m$ has a simple form
\begin{eqnarray} \label{D-psi}
&&D_-\psi_m(x)={\sqrt{2m}\over{L_H}}\psi_{m-1}(x), \cr
&&D_+\psi_m(x)=-{\sqrt{2(m+1)}\over{L_H}}\psi_{m+1}(x).
\end{eqnarray}

The equation (\ref{Del-mat-1}) in the notation $D_+$, $D_-$ has the form:
\begin{eqnarray} \label{eqnD+D-}
&&\Delta_n({\mathbf R})=V_0\rho_0\sum_{n'}\int d\widetilde{\bf R}{\cal K}_0(\widetilde{R})L_{nn'}(\widetilde{\bf R})\exp{\left(-{\widetilde{R}^2\over{4L_H^2}}\right)} \cr
&&\exp{\left({\widetilde{R}\over{2}}e^{-i\theta}D_+\right)}\exp{\left({\widetilde{R}\over{2}}e^{i\theta}D_-\right)}
\Delta_{n'}({\mathbf R}).
\end{eqnarray}
Here  we have used the formula
$e^{(\hat{P}+\hat{Q})}=e^{\hat{P}}e^{\hat{Q}}e^{-[\hat{P},\hat{Q}]/2}$.

The expressions (\ref{D-psi}) allow us to search for the solution of this equation in the form
\begin{equation} \label{Delta-x}
\Delta_{n}({\bf R}) = \Delta_{n} \psi_{n+m_0}(x).
\end{equation}
The constant $m_0$ corresponds to the dominant pairing channel that has the largest critical temperature $T_c$.

The coordinate dependence (\ref{Delta-x}) reduces the integral equation (\ref{eqnD+D-}) to a matrix equation.
The size of the matrix is formally infinite. However in a realistic situation one can easily apply a cut-off for the
size of the matrix. The order parameter $\Delta_n$ is always related to the Bessel function $J_n(k_F \rho_0)$. For realistic
$\rho_0$ these Bessel functions are small for large $n$. It allows us to neglect $\Delta_n$ with large $n$.

\subsection{n=0 dominant channel}

\begin{figure}
\includegraphics[width = 87mm, angle=-0]{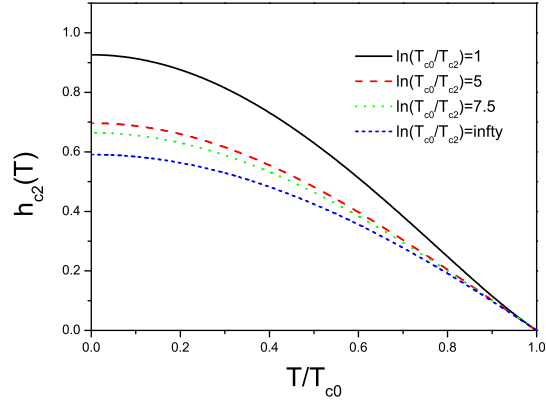}
\caption{Upper critical field $h_{c2}={H_{c2}(T)\over{T_{c0}dH_{c2}(T_{c0})/dT}}$ for different ratios of $T_{c0}/T_{c2}$ in the case when
the pairing in the channel $n=0$ is dominant.
}
\label{fig-clean-0}
\end{figure}

Let us first consider the case when the $n=0$ channel is dominant. It means that $T_{c0}>T_{cn}$ for $n \ne 0$. On the other hand $T_{cn}$ for $n \ne 0$ should be large enough in order to make relatively large effect on $H_{c2}(T)$.This situation takes place when $k_F\rho_0\approx 2$. In this case it is
sufficient to study the admixture of the order parameter $\Delta_2$ in the channel $n=2$.

 The dominant $n=0$ channel corresponds to $m_0 = 0$. The order parameter $\Delta_0$ has the coordinate dependence $\psi_0$ and the order parameter $\Delta_2$ has the coordinate dependence $\psi_2$. This dependence leads  to the matrix equation
\begin{equation} \label{meq-02}
\left(
\begin{array}{cc}
I_1 L_{00}& -I_3 L_{02} \\
-I_{3} L_{20} & I_2 L_{22}
\end{array}
\right)
\left(
\begin{array}{c}
\Delta_0 \\ \Delta_2
\end{array}
\right) = 0,
\end{equation}
where
\begin{equation}
\begin{array}{l}
I_1=\ln{{T\over{T_{c0}}}}+\int_0^{\infty}dx{1-\exp{(-{x^2\over{4z}})}\over{\sinh{(x)}}}, \\
I_2=\ln{{T\over{T_{c2}}}}+\int_0^{\infty}dx{1-\exp{(-{x^2\over{4z}})(1-{x^2\over{z}}+ {x^4\over{8z^2}})}\over{\sinh{(x)}}}, \\
I_3=\int_0^{\infty}dx{x^2\over{2\sqrt{2}z}}{\exp{(-{x^2\over{4z}})}\over{\sinh{(x)}}},
\end{array}
\end{equation}
$z={2\pi T^2\phi_0\over{v_F^2H}}$, $L_{00}=J_0^2(k_F\rho_0)$, $L_{22}=J_2^2(k_F\rho_0)$,
$L_{02}=L_{20}=J_0(k_F\rho_0)J_2(k_F\rho_0)$. Eq.(\ref{meq-02}) has a solution if $I_1I_2-I_3^2=0$.
Therefore all the details about the pairing potential are dropped out the equation for $H_{c2}$. The only information about the potential remains in the critical temperature in the channel with $n=0$, $T_{c0}$ and in the channel with $n=2$, $T_{c2}$

To calculate the critical field from the equation (\ref{meq-02}) one should find the maximal magnetic field $H$ when the equation has
 a non-trivial solution at the given temperature. The calculated critical fields for this case are presented on Fig.1. The main effect due to admixture
of the component of the gap with $n=2$ is the increase of the critical field up to 50\%. Moreover
the upward curvature of the temperature dependence $H_{c2}(T)$ is also clearly pronounced.

\subsection{n=2 dominant channel}

\begin{figure}
\includegraphics[width = 87mm, angle=-0]{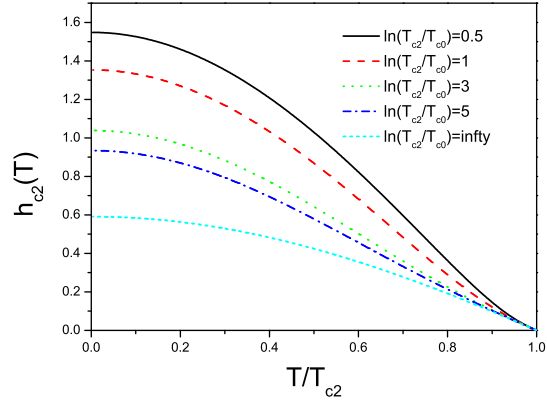}
\caption{Upper critical field $h_{c2}={H_{c2}(T)\over{T_{c2}dH_{c2}(T_{c2})/dT}}$ for different ratios of $T_{c2}/T_{c0}$ in the case when
the pairing in the channel $n=2$ is dominant.
}
\end{figure}

When the parameter $k_F\rho_0$  increases $k_F\rho_0\approx 3.5$, $T_{c2}>T_{c0}$ and the dominant channel is the one with
$n=2$. In this situation one should chose $m_0 = 2$ and the magnetic field
leads to the coupling between the channels with $n=-2,0,2$. The system
of equations for the order parameter reads:
\begin{equation} \label{meq-202}
\left(
\begin{array}{ccc}
I_{11}L_{22} & -I_{12}L_{20}& -I_{13}L_{22} \\
-I_{21}L_{02} & I_{22}L_{00} & -I_{23}L_{02} \\
-I_{31} L_{22} & -I_{32} L_{20} & I_{33} L_{22}
\end{array}
\right)
\left(
\begin{array}{l}
\Delta_{-2} \\ \Delta_0 \\ \Delta_2
\end{array}
\right) = 0,
\end{equation}
where
\begin{equation}
\begin{array}{l}
I_{11}=\ln{{T\over{T_{c2}}}}+\int_0^{\infty}dx{1-\exp{(-{x^2\over{4z}})}\over{\sinh{(x)}}}, \\
I_{22}=\ln{{T\over{T_{c0}}}}+\int_0^{\infty}dx{1-\exp{(-{x^2\over{4z}})(1-{x^2\over{z}}+ {x^4\over{8z^2}})}\over{\sinh{(x)}}}, \\
I_{33} = \ln{{T\over{T_{c2}}}}+\int_0^{\infty}dx{1-\exp{(-{x^2\over{4z}})(1-{2x^2\over{z}}+ {3x^4\over{4z^2}}-{x^6\over{12z^3}}+{x^8\over{384z^4}})}\over{\sinh{(x)}}},\\
I_{12} = I_{21} = \int_0^{\infty}dx{x^2\over{2\sqrt{2}z}}{\exp{(-{x^2\over{4z}})}\over{\sinh{(x)}}},\\
I_{13}=I_{31}=\int_0^{\infty}dx{\sqrt{6}x^4\over{48z^2}}{\exp{(-{x^2\over{4z}})}
\over{\sinh{(x)}}}\\
I_{23}=I_{32}={\sqrt{3}\over{2}}\int_0^{\infty}dx({x^2\over{z}}-{x^4\over{3z^2}}
+{x^6\over{48z^3}}){\exp{(-{x^2\over{4z}})}\over{\sinh{(x)}}}.
\end{array}
\end{equation}

Again the upper critical field is determined from the equation $Det(\hat I)=0$ ,where matrix
$\hat I$ is the left-hand side matrix in Eq. (\ref{meq-202}). All the details of the potential
are hidden into critical temperatures $T_{c0}$ and $T_{c2}$.

Fig.2 represents upper critical field calculated for this case. As it can be seen from this
picture there is a strong enhancement of the critical field $H_{c2}(0)$. Usually $H_{c2}(0)$
is expressed  via the slope of the critical field at $T_c$ (the orbital limiting field)
\cite{Ketterson}:
\begin{equation}
H_{c2}(0)=0.69T_c{dH_{c2}(T)\over{dT}}|_{T=T_c}.
\end{equation}
In the considered case $H_{c2}(0)$ is strongly enhanced in comparison with the orbital limiting field Eq.(22). The upward curvature of $H_{c2}(T)$ is even more pronounced than in Fig.1.

\section{Superconductor with impurities}\label{sec4}

In order to describe superconducting pairing in the presence of impurities we need to average
the product of two Green's functions in Eq. (\ref{delta2}) over impurities.
This averaging may be described
by the diagrammatic equation (Fig.3).  As it is presented in Fig. 3 the corresponding integral
equation contains single particles Green's functions averaged over impurities as well as
renormalized vertex\cite{AGD,Rickayzen}
\begin{eqnarray} \label{eq0}
&&K_\omega({\mathbf p},{\mathbf p'},{\mathbf k},{\mathbf k'})=K^0_\omega({\mathbf p},{\mathbf p'},{\mathbf
k},{\mathbf k'})+{n\over{(2\pi)^6}}\int d{\mathbf q}d{\mathbf l}d{\mathbf m}\cr
&&K^0_\omega({\mathbf p},{\mathbf l},{\mathbf k},{\mathbf m})|u({\mathbf q})|^2
K({\mathbf l}+{\mathbf q},{\mathbf p'},{\mathbf m}-{\mathbf q},{\mathbf k'}).
\end{eqnarray}
\begin{figure}
\includegraphics[width = 27mm, angle=-90]{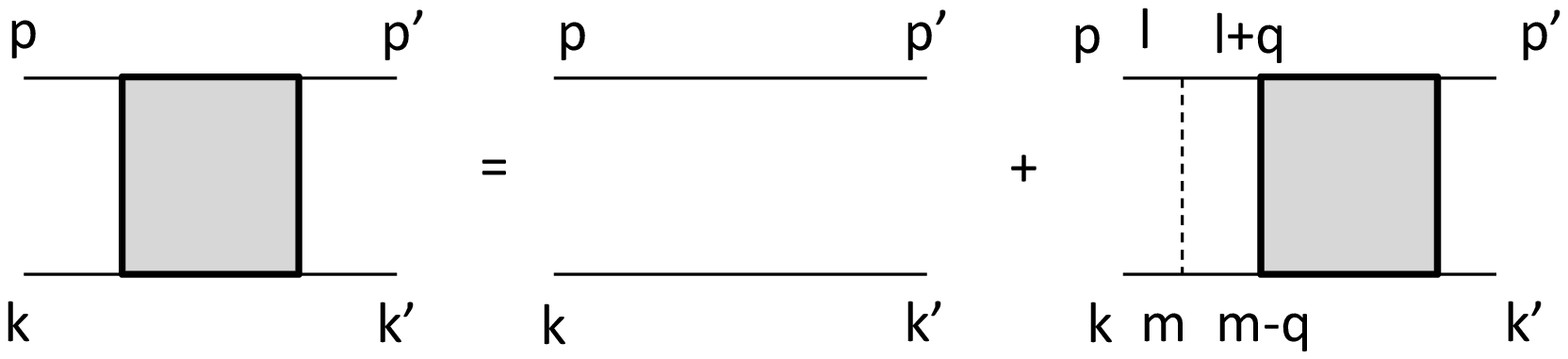}
\caption{Diagrammatic representation for the kernel $K({\mathbf p},{\mathbf p'},{\mathbf k},{\mathbf k'})$, where dashed line represents impurity scattering.}
\end{figure}
Here $K^0_\omega({\mathbf p},{\mathbf p'},{\mathbf k},{\mathbf k'}) = \overline{G}_\omega({\mathbf p},{\mathbf p'}) \overline{G}_{-\omega}({\mathbf k},{\mathbf k'})$ is the product of the two Green's functions $\overline{G}_\omega$ averaged over impurities separately.
$K_\omega({\mathbf p},{\mathbf p'},{\mathbf k},{\mathbf k'}) = \overline{G_{\omega}({\mathbf p},{\mathbf p'})G_{-\omega}({\mathbf k},{\mathbf k'})}$ is the same product where the Green's functions are averaged together. It corresponds to a term of the exact kernel of the equation (\ref{delta2}) related to the Matsubara frequency $\omega_n$. The exact kernel $K$ is the sum over the Matsubara frequencies $K = \sum_{\omega_n} K_{\omega_n}$.

Taking into account that the impurity scattering potential does not depend on the transmitted momentum  ${\mathbf q}$,
 equation (\ref{eq0}) can be solved using the coordinate representation. The details of this solution are described in the supplemental materials. The kernel $K_{\omega_n}$
can be expressed as a matrix in the basis $\psi_m$ corresponding to the macroscopic coordinates ${\bf R}$ and in the basis
$e^{in\varphi}$ for the internal coordinate ${\boldsymbol \rho}$. It is the same basis that we applied in the clean limit. Similarly to the
clean limit the values $n$ and $m$ are related $m=n+m_0$ where $m_0$ corresponds to the main order parameter.
\begin{eqnarray} \label{Knn2}
&&K_{n,n';\omega_n}=K_{n,n';\omega_n}^0 +{1\over{2\pi \tau N(0)}}
K^0_{n,\emptyset;\omega_n}K^0_{\emptyset,n';\omega_n}  + \cr
&&{1\over{(2\pi \tau N(0))^2}}
K^0_{n, \emptyset;\omega_n} K_{\emptyset,\emptyset;\omega_n}
K^0_{\emptyset,n';\omega_n},
\end{eqnarray}
\begin{equation} \label{K002}
K_{\emptyset,\emptyset;\omega_n}=\frac{K_{\emptyset,\emptyset;\omega_n}^0}
{1-(2\pi\tau N(0))^{-1}K_{\emptyset,\emptyset;\omega_n}^0},
\end{equation}
where $\tau$ is the impurity scattering time and $N(0)=m^*/2\pi$ is 2D density of states.
Here $\emptyset$ is the additional lower index corresponding to $n=0$, $\rho_0=0$. It appears due to the impurity scattering. The zero-order matrix elements $K_{n,n';\omega_n}^0$ can be expressed as integrals
\begin{eqnarray} \label{Knn0}
&& K_{n,n';\omega_n}^0 = \int \psi_{n+m_0} ({\bf R}) \psi_{n'+m_0} ({\bf R}') e^{in\varphi - in'\varphi'}  \cr
&& \overline{G}_{\omega_n}({\mathbf R}+{\boldsymbol
\rho}/2, {\mathbf R'}+{\boldsymbol \rho'}/2)\overline{G}_{-\omega_n}({\mathbf R}-{\boldsymbol
\rho}/2, {\mathbf R'}-{\boldsymbol \rho'}/2)  \cr
&& \delta(|{\boldsymbol \rho}| - \rho_0) \delta(|{\boldsymbol \rho}'| - \rho_0) d{\boldsymbol \rho} d{\boldsymbol \rho}' d {\bf R} d {\bf R}'.
\end{eqnarray}
The value of $\rho_0$ should be considered as $\rho_0 = 0$ when calculating $K_{n,n';\omega_n}^0$ with the lower index $\emptyset$ with equation (\ref{Knn0}).

The integral equation for the order parameter is reduced to the matrix equation
\begin{equation}\label{dirty-matrix}
\Delta_n=TV_0 \sum_{n',\omega_n}K_{n,n';\omega_n}
\Delta_{n'}.
\end{equation}

\subsection{n=0 dominant channel}

When the dominant channel of pairing corresponds to $n=0$ we apply $m_0=0$. It allows us to link indexes $m$ and $n$. After that using
Eqs.(\ref{Knn2},\ref{K002},\ref{Knn0}) we can calculate matrix $K_{n,n'} = \sum_{\omega} K_{n,n';\omega}$ for $n,n'=0,2$. The equation for $\Delta_0,\Delta_2$ has the form:
\begin{equation} \label{dir-mat-0}
\left(
\begin{array}{cc}
{\cal C}_{0,0} L_{00}& -{\cal C}_{0,2} L_{02} \\
-{\cal C}_{2,0} L_{20} & {\cal C}_{2,2} L_{22}
\end{array}
\right)
\left(
\begin{array}{c}
\Delta_0 \\ \Delta_2
\end{array}
\right) = 0
\end{equation}
Here the coefficients ${\cal C}_{n,n'}$ are
\begin{eqnarray}
&&{\cal C}_{0,0}=\ln{{T\over{T_{c0}}}}-2\sum_{n=0}^{\infty}\left[
{ll_HtI_1(y_n)\over{l-l_HI_1(y_n)}}-
{1\over{2n+1}}\right] \cr
&&{\cal C}_{2,0}={\cal C}_{0,2}=\sum_{n=0}^{\infty}{l_Ht\over{2\sqrt{2}}}{lI_2(y_n)
\over{l-l_HI_1(y_n)}}\cr
&&{\cal C}_{2,2}=\ln{{T\over{T_{c2}}}}-  2\sum_{n=0}^{\infty}\Bigl[l_HtI_3(y_n) + \cr &&
+
{l_H^2t\over{8}}{I_2^2\over{l-l_HI_1(y_n)}}-{1\over{2n+1}}\Bigr],
\end{eqnarray}
and
\begin{equation}
\begin{array}{l}
I_1(y)={1\over{y}}\int_0^{\infty}dx\exp{(-x-(x/2y)^2)} ,\\
I_2(y)={1\over{y^3}}\int_0^{\infty}dxx^2\exp{(-x-(x/2y)^2)}, \\
I_3(y)={1\over{y}}\int_0^{\infty}dx\exp{(-x-(x/2y)^2)}(1-{x^2\over{y^2}}+{x^4\over{8y^4}}).
\end{array}
\end{equation}
$y_n=l_H(lt(2n+1)+1)/l$, $l$ is the mean free path measured in units of
 $L_{T_{c0}}=v_F/2\pi T_{c0}=0.882\xi$, $l_H=L_H/L_{T_{c0}}$, and $t=T/T_{c0}$.

The upper critical field $H_{c2}$ corresponds to the largest magnetic field when the equation (\ref{dir-mat-0})
has a non-trivial solution.
Again all the details about the pairing potential are dropped out the equation for $H_{c2}$.
The only information about the potential remains in critical
temperatures $T_{c0}$ and $T_{c2}$.

\begin{figure}
\includegraphics[width=0.7\linewidth]{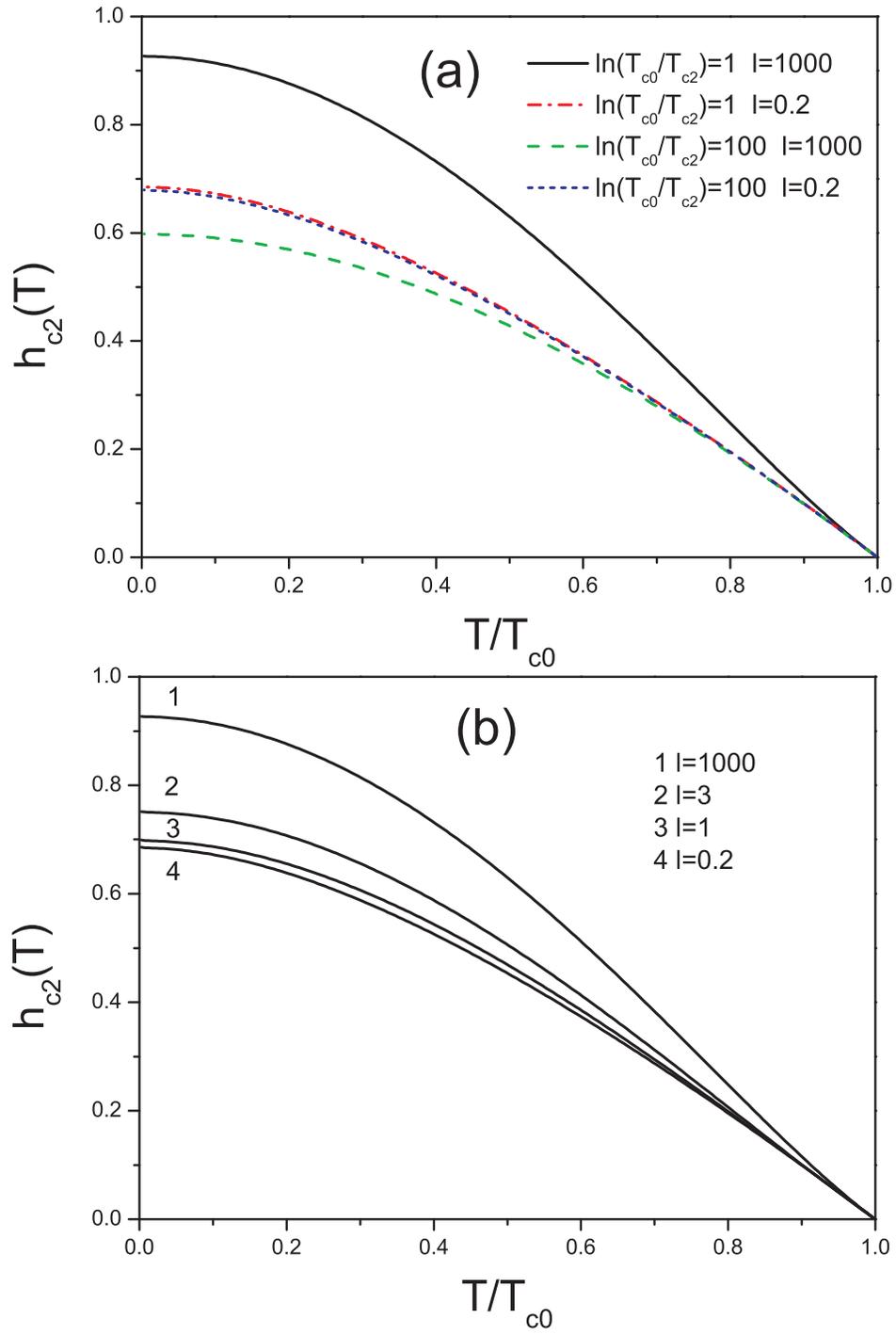}
\caption{(a) upper critical field for different ratios of $T_{c0}/T_{c2}$ and different mean free path of electrons $l$. (b) the dependence of the upper critical field on $l$ for $\ln(T_{c0}/T_{c2}) = 1$.  \label{fig:imp} }
\end{figure}

The temperature dependence of the upper critical field $H_{c2}(T)$ for different values of
the mean free path $l$ and different values of $\ln{(T_{c0}/T_{c2})}$ is presented in Fig. \ref{fig:imp}.
The upper critical field is normalized to the extrapolation of the linear $H_{c2}(T)$ dependence near $T_c$:
$h_{c2}={H_{c2}(T)/(T_{c0}dH_{c2}(T_{c0})/dT)}$ .
As it can be seen from Fig. \ref{fig:imp} (a) there is a big difference in $H_{c2}$ between curves with
$\ln{(T_{c0}/T_{c2})}=1$ and $\ln{(T_{c0}/T_{c2})}=100$ in the clean limit $l=1000$.
On the other hand these curves are almost indistinguishable in the dirty limit with $l=0.2$.
Note that the curves in the dirty limit are below the orbital limiting field (Eq.(22)).
On the other hand the curve of $H_{c2}(T)$ in the clean limit and for
$\ln{(T_{c0}/T_{c2})}=1$ exceeds substantially the orbital limiting field. Therefore we can
conclude that finite radius of interaction substantially increases the critical field only
in the clean limit. In the dirty limit the critical temperature in the channels with
$n \ne 0$ is strongly suppressed by the impurity scattering leading to suppression of the upper
critical field. The figure \ref{fig:imp}(b) illustrates this suppression for  finite values of $l$.
The effect is significantly suppressed at $l \lesssim 3$.

\section{Discussion}

The finite radius of the electron-electron interaction allows pairing into Cooper pairs with finite angular momentum $\Delta_n$. Each channel of pairing corresponds to its own critical temperature $T_{cn}$. However even when the temperature is larger than $T_{cn}$ of some non-dominant channel of the pairing, its order parameter can be generated in the presence of the external magnetic field. This generation is still related to the possibility of the existence of a given order parameter. For example when the leading parameter corresponds to $n=0$ and $T_{c2}$ tends to zero, $\ln(T_{c0}/T_{c2}) \rightarrow \infty$ and the effects of the coupling of $\Delta_0$ and $\Delta_2$ are absent (see Fig. \ref{fig-clean-0}).

The possibility of the existence of the order parameters with $n \ne 0$ is closely related to the conservation of the internal angular momentum of  Cooper pairs. In the dirty limit the impurity scattering is much stronger than electron-electron attraction. The angular momentum of the Cooper pair is therefore quickly lost due to scattering. It suppresses not only the critical temperature of the non-trivial order parameters but also the generation of these parameters in the magnetic field.

\begin{figure}
\includegraphics[width = 87mm, angle=-0]{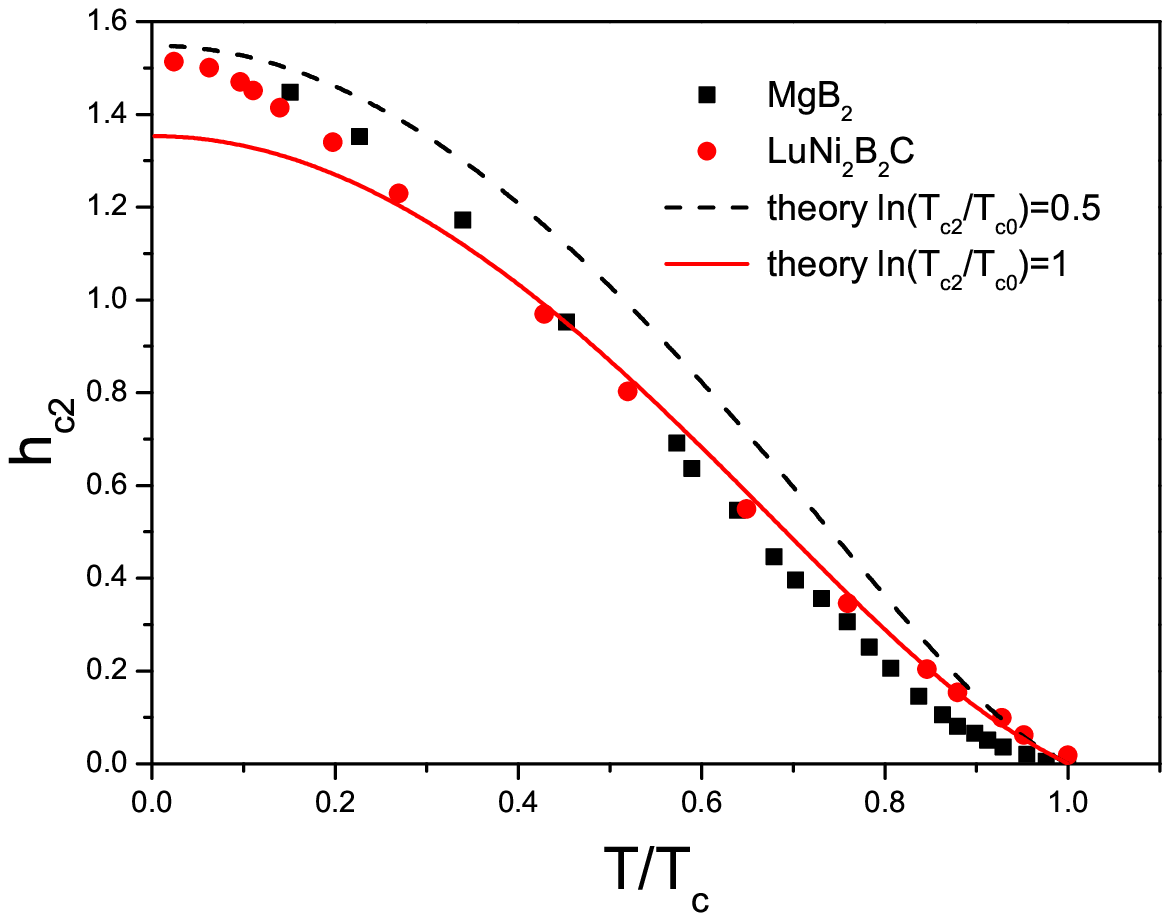}
\caption{Comparison of the theoretical model with the experiments on $MgB_2$\cite{Lyard} and on $LuNi_2B_2C$\cite{Shulga}}
\label{compar}
\end{figure}

The temperature dependence of $H_{c2}$ similar to the one that results from our theory was observed in $MgB_2$ \cite{Lyard} and in $LuNi_2 B_2C$ \cite{Shulga}.  In Fig. \ref{compar} we compare our theory with these experiments. The theory describes the experimental data relatively well. The small discrepancies can be attributed to the pure 2D character of the theory. We have discussed that the microscopic structure of the potential is reduced to the critical temperatures in different channels in terms of our theory.
For most of the relevant values of $k_F \rho_0$ only two channels can have relatively high $T_{cn}$ and can effectively affect the $H_{c2}(T)$ dependence with the selected coordinate form of the potential.
Therefore the present theory can describe only the situations when the $H_{c2}(T)$ dependence is governed by the interaction of two channels.
However, more complex forms of the pairing potentials can allow three or more channels with relatively high $T_{cn}$. We believe that such potentials can lead to further increase of the upper critical magnetic field at low temperatures.

In conclusion we generalize Werthamer-Helfand-Hohenberg theory to the case of a nonlocal
interaction between electrons. The theory is defined by the single
parameter $k_F\rho_0$. We show that when $k_F\rho_0 \gtrsim 1 $ the presence of the external magnetic
field leads to the generation of additional components of the order parameter with
different angular momentum. As a result the upper critical field is enhanced above the orbital
limiting field. The upward curvature in the temperature dependence of $H_{c2}(T)$ in the clean limit is predicted. The impurity scattering suppresses the effect in the dirty limit.

\vspace{3cm}

\hspace{6cm} \textbf{Supplemental materials}
\vspace{1cm}

\hspace{5cm} \textbf{Averaging of the kernel over impurities.}
\vspace{2cm}

In the main text we showed that the upper critical field $H_{c2}$ in a superconductor with impurities and a non-local pairing potential can be found from the integral equation (see eq. (5) from the main text). The kernel of this equation $K = \sum K_{\omega_n}$ is a sum over Matsubara frequencies of  products of two Green's functions averaged over impurities $K_{\omega_n} = \overline{G_{\omega_n}G_{-\omega_n}}$. Following the standard diagram rules for the impurity scattering \cite{AGD} the terms $K_{\omega_n}$ of the kernel themselves can be considered as a solution of the diagrammatic equation (see Fig. 3 and eq. (23) from the main text)
\begin{eqnarray} \label{eq0}
&&K_{\omega_n}({\mathbf p},{\mathbf p'},{\mathbf k},{\mathbf k'})=K^0_{\omega_n}({\mathbf p},{\mathbf p'},{\mathbf
k},{\mathbf k'})+{n\over{(2\pi)^6}}\int d{\mathbf q}d{\mathbf l}d{\mathbf m}
K^0_{\omega_n}({\mathbf p},{\mathbf l},{\mathbf k},{\mathbf m})|u({\mathbf q})|^2
K({\mathbf l}+{\mathbf q},{\mathbf p'},{\mathbf m}-{\mathbf q},{\mathbf k'})
\end{eqnarray}
Here $K^0_{\omega_n}({\mathbf p},{\mathbf p'},{\mathbf k},{\mathbf k'}) = \overline{G}_{\omega_n}({\mathbf p},{\mathbf p'}) \overline{G}_{-\omega_n}({\mathbf k},{\mathbf k'})$ is the term corresponding to the Matsubara frequency ${\omega_n}$ of the so-called zero-order kernel. It is equal to the product of the two Green's functions $\overline{G}_{\omega_n}$ averaged over impurities separately.
$K_{\omega_n}({\mathbf p},{\mathbf p'},{\mathbf k},{\mathbf k'}) = \overline{G_{\omega_n}({\mathbf p},{\mathbf p'})G_{-\omega_n}({\mathbf k},{\mathbf k'})}$ is the corresponding term of the exact kernel.

\begin{figure}[htbp]
    \centering
        \includegraphics[width=0.8\textwidth]{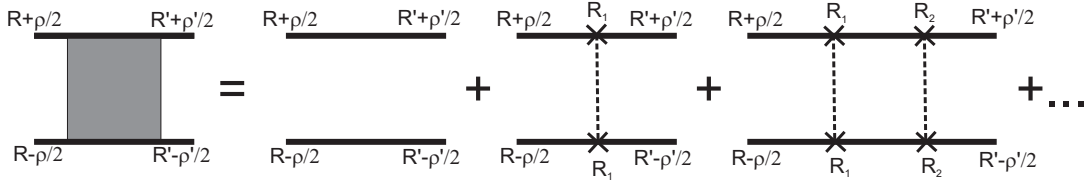}
        \caption{The equation for the kernel $K_{\omega_n}$ in the coordinate representation.  }
    \label{fig:ladder-R}
\end{figure}

In the present work we restrict ourselves with impurities with the scattering potential $u(q) = const$ independent on the scattering angle. This potential leads to the following equation for the kernel in the coordinate representation.
\begin{eqnarray}\label{eqK-R}
&&K_{\omega_n}({\mathbf R},{\mathbf R'},{\boldsymbol \rho},{\boldsymbol \rho'})=K^0_{\omega_n}({\mathbf R},
{\mathbf R'},{\boldsymbol \rho},{\boldsymbol \rho'}) +{1\over{2\pi N(0)\tau}} \int
d{\mathbf r}
K^0_{\omega_n}({\mathbf R},{\mathbf r},{\boldsymbol \rho},0) K_{\omega_n}({\mathbf r},{\mathbf R'},0,
{\boldsymbol \rho'})
\end{eqnarray}
Here $\tau$ is the impurity scattering time, $N(0)={m^*\over{2\pi}}$ is 2D density of states and
$m^*$ is the effective mass of electron. The kernel $K_{\omega_n}({\mathbf R},{\mathbf R'},{\boldsymbol \rho},{\boldsymbol \rho'})$ in the
coordinate representation is the averaged product of the two Green's functions
\begin{equation}\label{K-R}
K_{\omega_n}({\mathbf R},{\mathbf R'},{\boldsymbol \rho},{\boldsymbol \rho'})= \overline{  G_{\omega_n}({\mathbf R}+{\boldsymbol
\rho}/2, {\mathbf R'}+{\boldsymbol \rho'}/2)G_{-\omega_n}({\mathbf R}-{\boldsymbol
\rho}/2, {\mathbf R'}-{\boldsymbol \rho'}/2)  }.
\end{equation}
The zeros-order kernel $K^0_{\omega_n}({\mathbf R},{\mathbf R'},{\boldsymbol \rho},{\boldsymbol \rho'})$ corresponds to the similar product where the two Green's functions are averaged separately.
\begin{equation}\label{K0-R}
K^0_{\omega_n}({\mathbf R},{\mathbf R'},{\boldsymbol \rho},{\boldsymbol \rho'})= \overline{G}_{\omega_n}({\mathbf R}+{\boldsymbol
\rho}/2, {\mathbf R'}+{\boldsymbol \rho'}/2)\overline{G}_{-\omega_n}({\mathbf R}-{\boldsymbol
\rho}/2, {\mathbf R'}-{\boldsymbol \rho'}/2)  .
\end{equation}
Here the averaged Green's function is
\begin{equation}\label{Gav}
\overline{G}_{\omega_n}({\bf r}_1, {\bf r}_2) = \exp(i\phi({\bf r}_1,{\bf r}_2))
\begin{cases}
{-im\over{\sqrt{2\pi k_F |{\bf r}_1 - {\bf r_2}|}}}e^{i(k_F |{\bf r}_1 - {\bf r_2}|-\pi/4)-({|\omega_n|\over{v_F}}+l)|{\bf r}_1 - {\bf r_2}|},& \omega_n>0, \\
{im\over{\sqrt{2\pi k_F |{\bf r}_1 - {\bf r_2}|}}}e^{-i(k_F |{\bf r}_1 - {\bf r_2}|-\pi/4)-({|\omega_n|\over{v_F}}+l)|{\bf r}_1 - {\bf r_2}|},& \omega_n<0.
\end{cases}
\end{equation}
$\phi({\bf r}_1,{\bf r}_2)$ is the phase appearing in the magnetic field. This phase is discussed in the main text.
$l$ is the mean free path. We considered $k_F|{\bf r}_1 - {\bf r}_2| \gg 1$ similarly to the clean limit.

The ending points of the Green's functions in (\ref{K-R},\ref{K0-R}) are shifted by the small vector ${\boldsymbol \rho}$. This vector is related to the non-local nature of the pairing potential $V({\bf r} - {\bf r}')$.
In contrast to the pairing potential the impurity scattering potential is considered to be local. Accordingly in the intermediate points of the ladder shown on the figure \ref{fig:ladder-R} both Green's functions corresponding to the upper and the lower lines end at the same coordinates $R_1$, $R_2$, etc. In terms of the kernels $K^0$ and $K$ it corresponds to ${\boldsymbol \rho}=0$ in the intermediate points of the ladder.

In the main text we expanded the order parameter $\Delta({\bf R}, {\boldsymbol \rho}) = \Delta({\bf R}; \rho_0, \varphi) = \sum_n \Delta_n({\bf R})e^{in\varphi}$ over the values of the angle moment of vector ${\boldsymbol \rho}$. It allows us to consider the kernel $K_{\omega_n}$ as a matrix $K_{nn';\omega_n}$.
\begin{eqnarray}
K_{n,n';\omega_n}({\mathbf R},{\mathbf R'})={1\over{(2\pi)^2}}\int d{\boldsymbol \rho}
d{\boldsymbol \rho'}
e^{in\varphi-in'\varphi'}
\delta(|{\boldsymbol \rho}|-\rho_0)\delta(|{\boldsymbol \rho'}|-\rho_0)
K_{\omega_n}({\mathbf R},{\mathbf R'},{\boldsymbol \rho},{\boldsymbol \rho'})
\end{eqnarray}
The possibility of non-zero $n$ is related to finite values of the Bessel functions $J_n(k_F\rho_0)$.

The equation (\ref{eqK-R}) also allows the matrix representation. However the intermediate points correspond to the impurity potential and to the effective value $\rho_0 =0$. Therefore alongside with the matrix elements $K_{n,n';\omega_n}$ the equation includes
\begin{equation}
\begin{array}{l}
K_{n,\emptyset;\omega_n}({\mathbf R},{\mathbf R'})={1\over{(2\pi)}}\int d{\boldsymbol \rho}
e^{in\varphi}
\delta(|{\boldsymbol \rho}|-\rho_0)
K_{\omega_n}({\mathbf R},{\mathbf R'},{\boldsymbol \rho},0) ,
\\
K_{\emptyset,n';\omega_n}({\mathbf R},{\mathbf R'})={1\over{(2\pi)}}\int
d{\boldsymbol \rho'}
e^{-in'\varphi'}
\delta(|{\boldsymbol \rho'}|-\rho_0)
K_{\omega_n}({\mathbf R},{\mathbf R'},0,{\boldsymbol \rho'}) , \quad {\rm and}
\\
K_{\emptyset,\emptyset;\omega_n}({\mathbf R},{\mathbf R'})=
K_{\omega_n}({\mathbf R},{\mathbf R'},0,0) .
\end{array}
\end{equation}
The resulting matrix equation reads
\begin{eqnarray} \label{Knn1}
&&K_{n,n';\omega_n}({\mathbf R},{\mathbf R'})=[K_{n,n';\omega_n}^0({\mathbf R},{\mathbf R'})+{1\over{2\pi \tau N(0)}}\int d{\mathbf r}
K^0_{n,\emptyset;\omega_n}({\mathbf R},{\mathbf r})K^0_{\emptyset,n';\omega_n}({\mathbf r},{\mathbf R'})  + \cr &&
+{1\over{(2\pi \tau N(0))^2}}\int d{\mathbf r}d{\mathbf r_1}
K^0_{n, \emptyset;\omega_n}({\mathbf R},{\mathbf r})  K_{\emptyset,\emptyset;\omega_n}({\mathbf r},{\mathbf r_1})
K^0_{\emptyset,n';\omega_n}({\mathbf r_1},{\mathbf R'}),
\end{eqnarray}
where the term of the kernel $K_{\emptyset,\emptyset;\omega_n}({\mathbf R},{\mathbf R'})$ satisfy the equation:
\begin{equation}\label{K001}
K_{\emptyset,\emptyset;\omega_n}({\mathbf R},{\mathbf R'})=K^0_{\emptyset,\emptyset;\omega_n}({\mathbf R},{\mathbf R'})+
{1\over{2\pi \tau N(0)}}\int d{\mathbf r}K^0_{\emptyset,\emptyset;\omega_n}({\mathbf R},{\mathbf r})
K_{\emptyset,\emptyset;\omega_n}({\mathbf r},{\mathbf R'}).
\end{equation}

In the equation (\ref{Knn1}) we substituted the dependence on the coordinate ${\boldsymbol \rho}$ with the matrix. The next step is to reduce the coordinate ${\bf R}$ to the matrix elements. To do it we choose the basis functions $\psi_m(x)$ defined in Eq.(14) in the main text. Then we consider the kernels $K_{\omega_n}$ and $K_{\omega_n}^0$ as operators. For example, the Kernel $K_{nn';\omega_n}({\bf R}, {\bf R'})$ corresponds to the operator
\begin{equation}
\widehat{K}_{nn';\omega_n} f({\bf R}) = \int d {\bf R}' K_{nn';\omega_n}({\bf R}, {\bf R'}) f({\bf R'}).
\end{equation}
These operators can be represented with matrixes corresponding to the basis $\psi_m$. It appears that analogously to the clean limit discussed in the main text the operators $\widehat{K}_{nn';\omega_n}$ allow direct relation between indexes $n$ and $m$: $n=m-m_0$. The value $m_0$ corresponds to the leading order parameter. The indexes $\emptyset$ should be considered as $n=0$ for the purpose of this relation. The equations (\ref{Knn1},\ref{K001}) in this basis are reduced to
\begin{equation} \label{Knn2}
K_{n,n';\omega_n}=K_{n,n';\omega_n}^0 +{1\over{2\pi \tau N(0)}}
K^0_{n,\emptyset;\omega_n}K^0_{\emptyset,n';\omega_n}  +
{1\over{(2\pi \tau N(0))^2}}
K^0_{n, \emptyset;\omega_n} K_{\emptyset,\emptyset;\omega_n}
K^0_{\emptyset,n';\omega_n},
\end{equation}
\begin{equation} \label{K002}
K_{\emptyset,\emptyset;\omega_n}=\frac{K_{\emptyset,\emptyset;\omega_n}^0}
{1-(2\pi\tau N(0))^{-1}K_{\emptyset,\emptyset;\omega_n}^0}.
\end{equation}

The equations (\ref{Knn2},\ref{K002}) actually represent the solution of the equation (\ref{eq0}). Naturally, the matrix elements of zero-order kernel can be written explicitly
\begin{eqnarray}
&& K_{n,n';\omega_n}^0 = \frac{1}{(2\pi)^2}  \int \psi_{n+m_0} ({\bf R}) \psi_{n'+m_0} ({\bf R}') e^{in\varphi - in'\varphi'} \times
\cr && \times \overline{G}_{\omega_n}({\mathbf R}+{\boldsymbol
\rho}/2, {\mathbf R'}+{\boldsymbol \rho'}/2)\overline{G}_{-\omega_n}({\mathbf R}-{\boldsymbol
\rho}/2, {\mathbf R'}-{\boldsymbol \rho'}/2)  d\varphi d\varphi' d {\bf R} d {\bf R}',
\end{eqnarray}
where the vectors ${\boldsymbol \rho}$ and ${\boldsymbol \rho}'$ have the absolute value $\rho_0$ and the polar angles $\varphi$ and $\varphi'$ correspondingly. The matrix elements of the kernel $K_{n,n';\omega_n}$ can be obtained directly from $K_{n,n';\omega_n}^0$ with equations (\ref{Knn2},\ref{K002}). The total kernel $K_{n,n'} = \sum_{\omega_n}K_{n,n';\omega_n}$ can be used to find the upper critical field as described in the main text.

\end{document}